\documentclass[a4paper, amsfonts, amssymb, amsmath, reprint, showkeys, footinbib, twoside,superscriptaddress,floatfix,longbibliography]{revtex4-1}

\usepackage{graphicx} 
\graphicspath{{./}}
\usepackage{amsmath}
\usepackage{comment}
\usepackage{booktabs}
\usepackage{hyperref}
\usepackage{url}
\usepackage{adjustbox}
\usepackage{array}
\usepackage[english]{babel}

\usepackage{multirow}
\usepackage{hhline}
\usepackage{makecell}
\usepackage[table]{xcolor}

\newcommand{\figLabelCapt}[1]{\textbf{\MakeLowercase{{#1}}}}
\newcommand{\refSub}[2]{\hyperref[#2]{\ref{#2}\figLabelCapt{#1}}}
\newcommand{\figref}[1]{Fig.~\ref{#1}}
\newcommand{\figrefsub}[2]{Fig.~\refSub{#1}{#2}}

\newcommand{\br}[1]{\mathbf{r}}
\newcommand{\bk}[1]{\mathbf{k}}

\begin{document}

\title{An equivariant classical density functional for ionic fluids at complex interfaces}

\author{Bingqing Cheng}
\email{bingqingcheng@berkeley.edu}
\affiliation{Department of Chemistry, UC Berkeley, California 94720, United States}
\affiliation{Chemical Sciences Division, Lawrence Berkeley National Laboratory, Berkeley, California, 94720, United States}
\affiliation{Bakar Institute of Digital Materials for the Planet, UC Berkeley, California 94720, United States}

\date{\today}

\begin{abstract}
\textbf{Abstract:}
Predicting the combined effects of ionic packing and long-range electrostatics at complex interfaces is technologically important, but remains challenging without atomistic simulations. Here we present a machine-learned classical density functional for ionic fluids that combines a symmetry-adapted neural free energy for short-range correlations with explicit long-range electrostatics. Trained on three-dimensional equilibrium densities and bulk correlations of the full-Coulomb restricted primitive model, a single functional transfers to charge-patterned particles and atomically structured electrodes. Around anisotropic charged solutes, it reproduces the ionic structure and mean forces from molecular dynamics, including reversals of the bare electrostatic force. At metal electrodes, self-consistent coupling to metal atoms captures the redistribution of ionic and electrode charge and the loading-dependent capacitance enhancement produced by steps and bumps. Combined with mobilities determined independently from bulk molecular dynamics, the same functional describes the main features of charging dynamics and resolves geometry-dependent spatial relaxation. This framework can be extended to more complex ionic fluids, enabling efficient exploration of how composition and geometry control structure and response.
\end{abstract}

\maketitle

\section{Introduction}

Ionic fluids are ubiquitous in physical and biological systems, and studying their behavior is a long-standing topic in theory and experiment~\cite{plechkova2007applications,hayes2015structure}.
Around a solute, the competition between electrostatics,
ionic packing and entropy determines solvation and solvent-mediated forces. 
Near an electrode, ions reorganize into an electric double layer (EDL)~\cite{Kondrat2023Nanoconfinement,Hartel2017EDL},
where correlations and finite ion
size can produce layering, overscreening and crowding beyond the
point-ion Poisson--Boltzmann description~\cite{Bazant2011Overscreening}.
A useful theory should connect these local structures and electric fields to
thermodynamic and dynamical observables across different geometries.

Classical density functional theory (cDFT) provides a statistical-mechanical
framework for this purpose~\cite{Evans1979,Evans2016Developments},
by describing a fluid through its one-body density fields.
For a single-component system, the equilibrium density $\rho(\mathbf r)$ in an external
potential $V^{\mathrm{ext}}(\mathbf r)$ minimizes the grand potential
\begin{equation}
\Omega[\rho]
=F_{\mathrm{id}}[\rho]+F_{\mathrm{exc}}[\rho]
+\int\!\mathrm d\mathbf r\,\rho(\mathbf r)
\bigl[V^{\mathrm{ext}}(\mathbf r)-\mu\bigr],
\label{eq:grand_potential}
\end{equation}
where $\mu$ is the chemical potential, $F_{\mathrm{id}}$ is the known
ideal-gas contribution, and $F_{\mathrm{exc}}$ is generally unknown and contains the effects of
interparticle interactions. 

The central difficulty is to construct an accurate approximation for $F_{\mathrm{exc}}$.
Fundamental measure theory provides accurate
descriptions for hard-sphere systems~\cite{Rosenfeld1989,Roth2010FMT}.
For ionic fluids, however, packing must be combined with electrostatic
correlations. Existing ionic functionals go beyond a simple mean-field
treatment and can reproduce substantial features of EDL structure,
but their accuracy depends on concentration, surface charge and ionic
valence~\cite{Hartel2017EDL,Voukadinova2018EDL}.
Refinements of analytical electrostatic functionals can improve
ionic layering in the primitive model~\cite{BultmannHartel2022Primitive},
but accurate interfacial profiles do not necessarily imply accurate
bulk correlations or their asymptotic decay~\cite{Cats2021Decay}.

Machine learning offers a data-driven route to this
problem~\cite{SimonOettel2026Review}. 
Free-energy functionals can be
learned from equilibrium density
profiles~\cite{LinOettel2019,Cats2021MLFreeEnergy} or constrained through
their second derivatives by bulk pair
correlations~\cite{Dijkman2025PairMatching}. 
Neural operators directly learn
potential--density maps~\cite{Pan2025NeuralOperators,Yang2025HighDimensional}.
Neural functional theory
learns the one-body direct correlation functional from grand-canonical
atomistic simulation
data~\cite{Sammuller2023Neural}. This approach has been extended to
phase coexistence~\cite{Sammuller2025Coexistence}, mixtures~\cite{Robitschko2025Mixtures}, and training without supplied chemical
potentials~\cite{Sammuller2026ChemicalPotential}.
Although much of this work uses planar density profiles, learning
inhomogeneities in more than one spatial dimension is also
possible~\cite{Glitsch2025HigherDimensions,Weimar2026Cubic}.
Recently, Equi-cDFT was introduced as a symmetry-adapted framework that learns the 
scalar free energy from fully three-dimensional equilibrium
density fields~\cite{cheng2026equivariant}.

Many machine-learned density functionals exploit locality, representing the one-body
direct correlation functional or the local free-energy contribution using
finite-range density
environments~\cite{Sammuller2023Neural,SammullerSchmidt2024LocalPair,cheng2026equivariant}.
This reduces the space of the learned representations,
and naturally allows system-size generalization.
For ionic systems, however, the
Coulomb contribution is intrinsically nonlocal and requires a long-range
treatment. Local molecular field theory
(LMFT) accounts for the averaged effects of long-ranged components using an effective field~\cite{RodgersWeeks2008LMF,ArcherEvans2013LMF}.
Bui and Cox recently combined LMFT with neural functional theory for
ionic and polar fluids~\cite{Bui2025Ionic,bui2026dielectrocapillarity}. They learn direct correlation
functionals from simulations of a short-range mimic fluid and restore
the long-range electrostatic contribution through LMFT.

Here, we introduce a long-range electrostatic extension to Equi-cDFT.
The short-range extensive scalar free energy is predicted from
symmetry-adapted local density descriptors,
and a global Coulomb term is added.
We learn directly from three-dimensional canonical equilibrium
densities generated with atomistic systems with full Coulomb interactions. 

The learned free energy also supplies the thermodynamic driving force
for density evolution. Dynamical density functional theory
connects equilibrium functionals to time-dependent density fields
through an adiabatic closure~\cite{MarconiTarazona1999DDFT,ArcherEvans2004DDFT},
and learned free-energy functionals have already been applied in this
setting~\cite{Ram2025DDFT}. Here we combine the learned mixture
functional with a mobility matrix obtained independently
from molecular dynamics (MD). This separates the equilibrium free-energy
model from the transport information.

We use the restricted primitive model (RPM):
oppositely charged, equal-sized ions interact in a uniform dielectric
background. After benchmarking the bulk thermodynamics and charge correlations (see Methods),
we examine solvation and solvent-mediated forces around charged nonspherical
particles, and EDL structure and charging at atomically structured
electrodes. 

\section{Theory}

\subsection{Long-range free-energy functional}

$\rho_a(\mathbf r)$ is the number-density field of species $a$
on a regular cubic grid with voxel volume $\Delta V = h^3$,
where $h$ is the evaluation grid spacing and $h_0$ denotes the training spacing.
Equi-cDFT represents the local environment $\chi_g$ of each grid point $g$ at $\mathbf r_g$ using
a modified Cartesian atomic cluster expansion
(CACE)~\cite{cheng2024cartesian,drautz2019atomic}:
we first apply a fixed smoothing operator $G$ to
the density, $\widetilde\rho_a=G\rho_a$, and then construct Cartesian moments
\begin{equation}
A_{an\boldsymbol\ell}(\mathbf r_g)
=\sum_{0<|\mathbf q|\leq q_{\mathrm{cut}}}
w_n(|\mathbf q|)\widetilde\rho_{a,g+\mathbf q}\,
q_x^{\ell_x}q_y^{\ell_y}q_z^{\ell_z},
\label{eq:lattice_cace_moments}
\end{equation}
where $\mathbf q=(q_x,q_y,q_z)$ denotes the
integer offsets of neighboring voxels within the cutoff
$q_{\mathrm{cut}}=r_{\mathrm{cut}}/h$, where $r_{\mathrm{cut}}$ is the physical cutoff,
$\boldsymbol\ell=(\ell_x,\ell_y,\ell_z)$ specifies the Cartesian powers,
and $w_n$ is a radial basis. 
Invariant features are obtained by summing products over symmetry operations
$\mathcal R$
in the cubic point group $O_h$,
\begin{equation}
B_{an;\boldsymbol\ell_1,\ldots,\boldsymbol\ell_\nu}(\mathbf r_g)
=
\sum_{\mathcal R\in O_h}
\prod_{m=1}^{\nu}
s_{\boldsymbol\ell_m}(\mathcal R)
A_{an,\mathcal R\boldsymbol\ell_m}(\mathbf r_g),
\label{eq:lattice_cace_invariants}
\end{equation}
where a moment transforms as
$A_{an\boldsymbol\ell}
\longrightarrow
s_{\boldsymbol\ell}(\mathcal R)
A_{an,\mathcal R\boldsymbol\ell}$ under operation $\mathcal R$.
The nonredundant $B$-features and central densities form $\chi_g$.
A neural network shared by all grid points maps $\chi_g$ and temperature $T$
to a local excess free energy per particle, $a_{\mathrm{exc},a}(\chi_g,T)$,
for each species.
The learned short-range contribution is
\begin{equation}
F^{\mathrm{SR}}_{\mathrm{exc}}[\{\rho_a\},T]
=\Delta V\sum_{g,a}\widetilde\rho_{a,g}\,a_{\mathrm{exc},a}(\chi_g,T).
\label{eq:local_energy_decomposition}
\end{equation}

For species with fixed valencies $z_a$, define the charge-number density
$\rho_Z(\mathbf r)=\sum_a z_a\rho_a(\mathbf r)$.
In a neutral periodic cell of volume $V$, the
long-range contribution is
\begin{equation}
F^{\mathrm{LR}}_{\mathrm{exc}}
=\frac{e^2}{2\epsilon_0\epsilon_r V}\sum_{\mathbf k\ne0}
\frac{e^{-\alpha k^2}}{k^2}
\left|\widehat\rho_Z(\mathbf k)\right|^2,
\label{eq:long_range_energy}
\end{equation}
where $\epsilon_0$ is the vacuum permittivity, $\epsilon_r$ is the relative
permittivity, $\alpha$ is a fixed parameter, and
$\widehat\rho_Z(\mathbf k)=\Delta V\sum_g\rho_{Z,g}e^{-i\mathbf k\cdot\mathbf r_g}$.

The total excess functional is
\begin{equation}
F_{\mathrm{exc}}=F^{\mathrm{SR}}_{\mathrm{exc}}+F^{\mathrm{LR}}_{\mathrm{exc}}.
\end{equation}
Both the short-range and long-range contributions are expressed through
representations of the density fields: local symmetry-adapted moments for the
former and global Fourier modes for the latter. With physical length
scales held fixed and the discretization adjusted consistently, the same
learned parameters can be used at different grid spacings.
This is closely related to the discretization-transfer principle of
neural-operator learning~\cite{Kovachki2023NeuralOperator,Li2021FourierOperator}.
We examine grid-spacing transfer
in Methods: the Equi-cDFT for the RPM was trained using grid spacing $h_0=0.25\sigma$, yet is transferable to $h=0.125\sigma$ and $0.5\sigma$.

The one-body direct correlation functional
\begin{equation}
c^{(1)}_{a,g}([\rho])
=-\frac{\beta}{\Delta V}
\frac{\partial F_{\mathrm{exc}}[\rho]}
{\partial\rho_{a,g}},
\label{eq:c1}
\end{equation}
can be computed by automatic differentiation, where $\beta=(k_{\mathrm B}T)^{-1}$.
The local chemical potential is
\begin{equation}
\beta \mu^{\mathrm{loc}}_{a,g}
=\ln\!\left(\Lambda_a^3\rho_{a,g}\right)
+\beta V^{\mathrm{ext}}_{a,g}
-c^{(1)}_{a,g}([\rho],T),
\label{eq:local_chemical_potential}
\end{equation}
where the first term is the ideal contribution, $\Lambda_a$ denotes the thermal wavelength of species $a$, and $V^{\mathrm{ext}}_{a,g}$ is its external
potential energy.
Equilibrium requires a spatially constant
chemical potential for each species.

When training on equilibrium density fields collected from
canonical MD under different species-dependent potentials $V^{\mathrm{ext}}$,
the chemical-potential balance loss is used:
\begin{equation}
\mathcal L^{\Delta\mu}
=\sum_{g,a}\left|
\beta \mu^{\mathrm{loc}}_{a,g}
-\overline{\beta \mu^{\mathrm{loc}}_a}
\right|^2,
\label{eq:chemical_potential_balance_loss}
\end{equation}
where the sum runs over the reliably sampled region and the overbar denotes
the spatial mean for each species.
Here we supplement the field-based loss with
bulk-response targets derived from MD structure factors~\cite{Dijkman2025PairMatching}:
Second functional derivatives
of $F_{\mathrm{exc}}$ determine the pair direct correlation functions,
which are related to bulk structure factors by the mixture
Ornstein--Zernike equation.
Equilibrium densities minimize the grand potential at fixed chemical
potentials, or the total free energy at fixed species particle numbers.

\subsection{Dynamical cDFT}

We use a deterministic, adiabatic dynamical cDFT for the mean density fields.
The same equilibrium functional supplies the instantaneous local chemical
potentials, while a mobility matrix couples the species fluxes:
\begin{equation}
\frac{\partial\rho_a}{\partial t}=-\nabla\cdot\mathbf J_a,
\qquad
\mathbf J_a=-\sum_b M^0_{ab}(\{\rho\})\nabla\mu_b^{\mathrm{loc}}.
\label{eq:ddft_flux}
\end{equation}
For a symmetric positive-semidefinite $\mathbf M^0$ and fixed external potentials,
this evolution decreases the free energy. 
This deterministic
evolution should be compared to MD ensemble-mean relaxation.

We make the approximation that $\mathbf M^0$ at grid point $g$ depends only on the local densities $\rho(\mathbf{r}_g)$ of all species.
We then estimate $\mathbf M (\rho)$ from independent bulk MD simulations. 
In MD, the number-density wave for species $a$ is $\widetilde{\rho}_a(\mathbf k,t)= (1/V)\sum_{j\in a}e^{-i\mathbf k\cdot\mathbf r_j(t)}$,
for $\mathbf k\ne0$.
The corresponding time-dependent correlation function is
\begin{equation}
C_{ab}(k,t)=\langle 
\widetilde{\rho}_a(\mathbf k,t)\widetilde{\rho}^*_b(\mathbf k,0)\rangle.
\end{equation}
Writing
\begin{equation}
\mathbf I(k)=\int_0^{\infty}\mathbf C(k,t)\,\mathrm dt,
\label{eq:Ik}
\end{equation}
the linear-response estimate is
\begin{equation}
\mathbf M(k)
=\frac{V}{k_{\mathrm B}T\,k^2}
\mathbf C(k,0)\,\mathbf I(k)^{-1}\,\mathbf C(k,0).
\label{eq:mobility_from_md}
\end{equation}
Extrapolation to small $k$, $\mathbf M^0 = \lim_{k \rightarrow 0} \mathbf M(k)$, provides the mobility used in
Eq.~\eqref{eq:ddft_flux}.

\section{Results}

\subsection{Interaction between charge-patterned particles}

\begin{figure*}[!t]
\centering
\includegraphics[width=\textwidth]{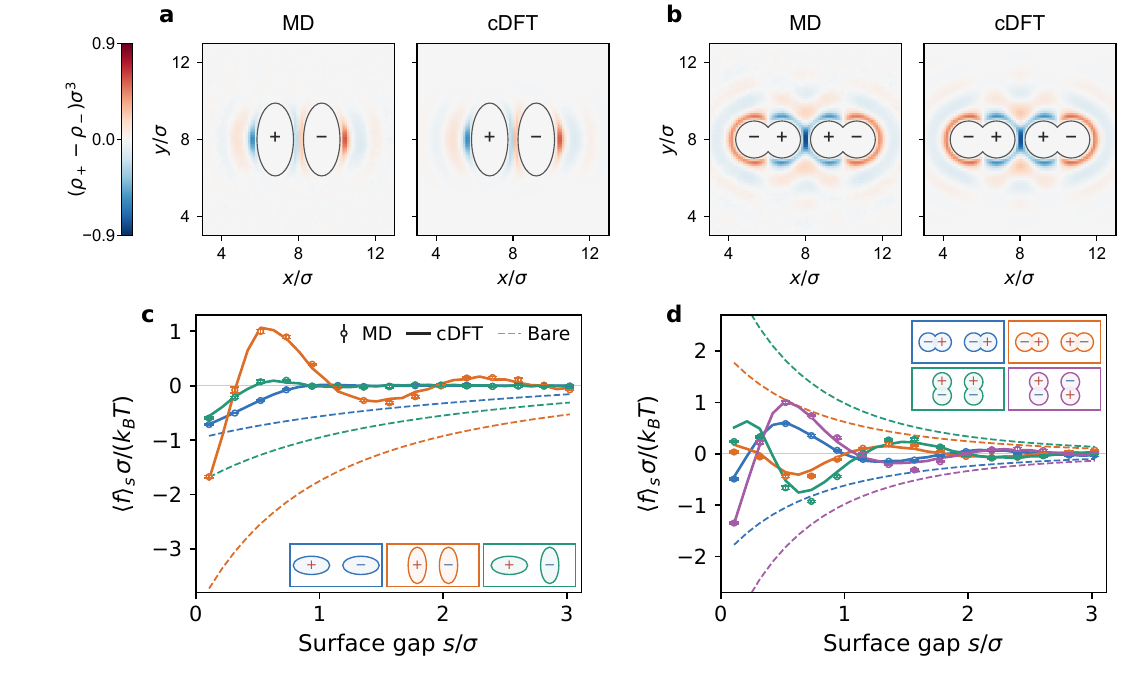}
\caption{\textbf{Ionic structure and forces around nonspherical solutes.}
\textbf{a,b} Ionic charge-number density $(\rho_+-\rho_-)\sigma^3$,
averaged over a $1\sigma$-thick slab through the solute-particle centers: side-by-side ellipsoids (\textbf{a}) and dumbbells
with positive ends facing (\textbf{b}). Outlines mark the $10k_{\mathrm B}T$
repulsive-core contours.
The surface gap for both is $s=0.5244\sigma$.
\textbf{c,d} Total mean separating force $\langle f\rangle_s$, versus surface gap $s$
for oppositely charged ellipsoids (\textbf{c}) and bipolar dumbbells
(\textbf{d}). Positive force denotes repulsion. MD standard statistical
uncertainties are indicated by the error bars.}
\label{fig:colloid}
\end{figure*}

Charged particles immersed in ionic media occur in dispersions,
polyelectrolyte solutions, and biomolecular assemblies. Their interactions
depend on their total charge, particle shape and the spatial arrangement
of charged groups~\cite{Bianchi2011InversePatchy,Lytle2019ChargeSequence}.
These features affect the three-dimensional ionic
structure, and in turn, the mediated interactions between the immersed particles.

Here we use two sets of geometries: oppositely charged prolate ellipsoids, and bipolar dumbbells. 
Changing separation and orientation
alters both the electrostatic environment and the space available to the ions.
For every fixed-particle configuration, we minimize the same learned functional in the periodic cubic cell of edge length $16\sigma$ and a $0.125\sigma$ grid, with RPM ionic particle numbers $N_+=N_-=820$.

The charge-density maps in \figrefsub{a}{fig:colloid} and
\figrefsub{b}{fig:colloid} reveal distinct screening patterns around the
two particle geometries. For the oppositely charged ellipsoids, the
screening layers are distributed asymmetrically between the facing and
outward surfaces. For the dumbbells with positive ends facing, negative
ionic charge accumulates in the gap, while positive
screening charge surrounds the outward negative ends. Alternating charge
layers extend into the surrounding fluid. Equi-cDFT reproduces these
spatial patterns, including the localized gap response and the broader
screening layers seen in MD.

Having established that the functional captures these ionic structures,
we next examine whether the equilibrium densities reproduce the effective
interactions between the particles. The classical Feynman--Hellmann
identity~\cite{Feynman1939Forces} gives the total mean separating force:
\begin{equation}
\langle f\rangle_s
=-\sum_a\int\!\mathrm d\mathbf r\,
\rho_{a,s}(\mathbf r)
\frac{\partial V^{\mathrm{ext}}_a(\mathbf r;s)}{\partial s}
-\frac{\mathrm d U_{\mathrm{bare}}(s)}{\mathrm ds},
\label{eq:colloid_feynman_hellmann_grid}
\end{equation}
where $s$ is the surface gap at fixed particle orientations, $V_a^{\mathrm{ext}}$
is the external potential acting on species $a$, and $\rho_{a,s}$ is its
equilibrium density. 
The integral is the contribution from the ionic fluid, and the last term is the bare force between the fixed
particle charges.

\figrefsub{c}{fig:colloid} and \figrefsub{d}{fig:colloid} show that the
total mean forces differ qualitatively from the dashed bare-force curves.
At the separation about $s=0.52$ shown in the density maps, side-by-side oppositely
charged ellipsoids repel despite their bare attraction, whereas bipolar
dumbbells with positive ends facing attract despite their bare repulsion.
In both cases, the ionic contribution opposes and
outweighs the bare force, reversing the tendency of the particles to
approach or separate.

Moreover, the total forces vary nonmonotonically with separation, with attractive
and repulsive intervals depend on particle
orientation. Changing orientation can therefore change whether particles
attract or repel at the same surface gap. These trends show that screening
cannot be represented here simply as a uniform attenuation of the bare
electrostatic force. 

The geometry-dependent mean forces suggest that the ionic medium can
influence both whether charged objects associate and how they make
contact. For anisotropic or charge-patterned particles, changing the
balance between end-to-end and side-by-side interactions could favor
chains, bundles or compact aggregates~\cite{Bianchi2011InversePatchy}.
The implications also extend to adsorption at charged
surfaces. Ionic rearrangement can promote or resist contact, while
separation-dependent forces can favor a finite
surface gap or create a barrier to closer
approach~\cite{Sidhu2018NanoparticleWall}. The force reversals observed
here illustrate how these tendencies can differ qualitatively from
those suggested by the bare charges. Particle shape and charge
pattern could therefore tune association behavior by organizing the nearby ionic fluid.

\subsection{Electric double layers and capacitance}

Electric double layers govern how charge is stored at an electrode--electrolyte
interface and are central to electrochemical
capacitors~\cite{Hartel2017EDL,Kondrat2023Nanoconfinement}.
Their response depends on the available surface area and on how ions pack
and screen electric fields near that surface. MD simulations
have shown that changing electrode topography can qualitatively alter
the capacitance through changes in interfacial ionic
structure~\cite{Vatamanu2011Topography}. At a metal surface this is a
coupled problem: steps and protrusions alter ionic packing, while the
metal charge redistributes in response to both the applied voltage and
the ions. 

\begin{figure*}[!t]
\centering
\includegraphics[width=\textwidth]{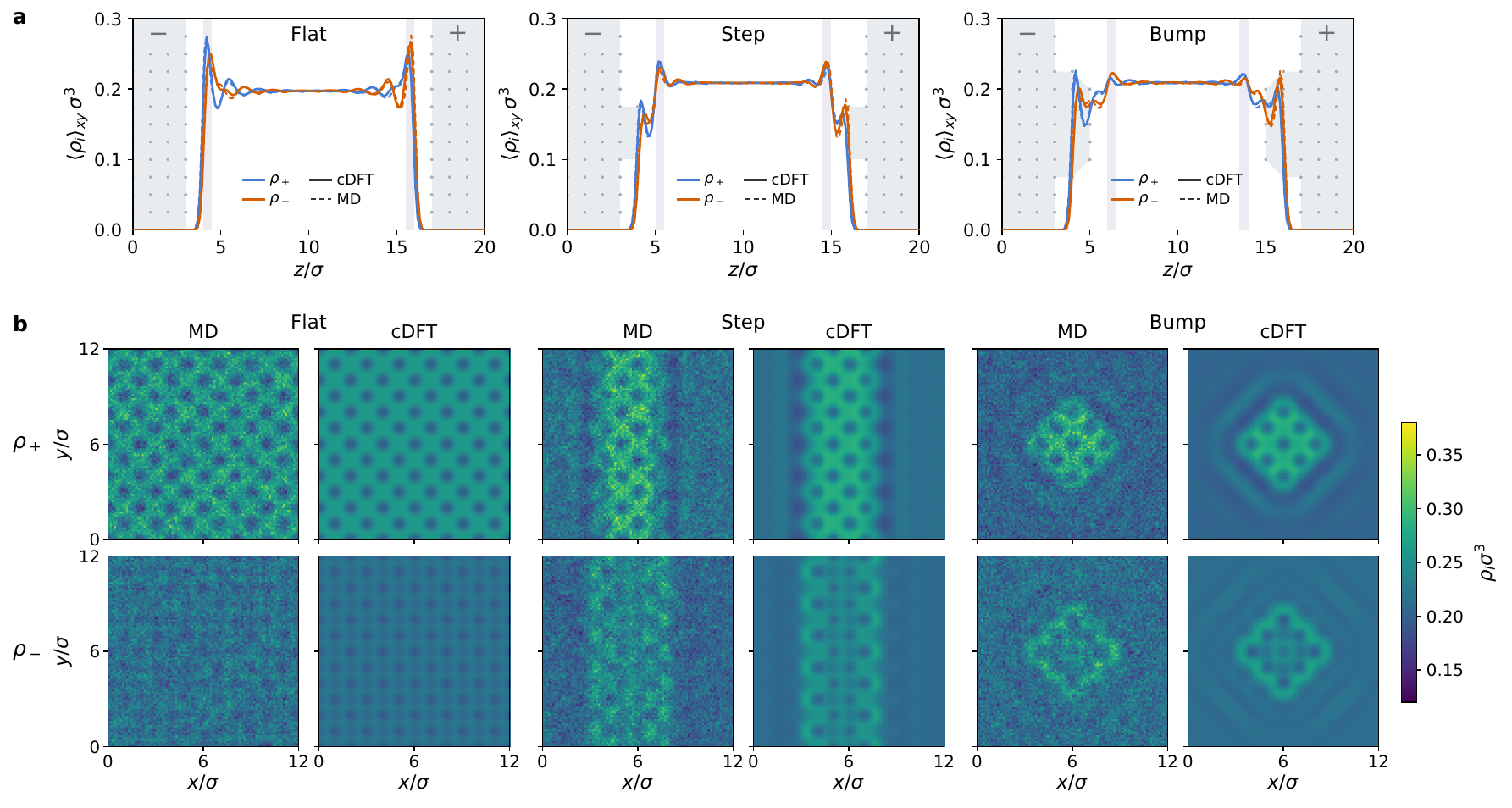}
\caption{\textbf{Equilibrium double layers at atomically structured metal electrodes, with $N_+=N_-=352$ of ion loading and at 2V of applied electric bias.}
\textbf{a} Planar-averaged species densities.
Narrow shaded bands show one MD standard error.
Gray sites and shapes show side views of the electrode
geometry. Purple bands mark the fixed $0.5\sigma$-thick first-layer
windows used in \textbf{b}.
\textbf{b} Lateral species-density maps from both MD and cDFT.}
\label{fig:edl-structure}
\end{figure*}

We examine flat surfaces and surfaces with steps or
bumps at matched ionic loading and voltage. In both MD and cDFT, explicit
metal atoms carry flexible Gaussian charges under the Siepmann--Sprik
scheme~\cite{siepmann1995influence}. The ionic densities and metal charges
are determined self-consistently under the imposed voltage.
We minimize the same learned fluid functional for all three geometries on a
$0.125\sigma$ grid in a periodic $12\sigma\times12\sigma\times20\sigma$
cell for $N_+=N_-=176,264,352,440$ and imposed voltages
$V=0,0.5,1,1.5,2\,\mathrm V$ across the full periodic cell.
\figref{fig:edl-structure} shows the density profiles and lateral maps
for $N_+=N_-=352$ at $V=2\,\mathrm V$. 

The planar profiles in \figrefsub{a}{fig:edl-structure} show ionic
packing and charge separation near the electrodes. Counterions
are enriched closest to each charged surface, while coions also retain a
structured interfacial profile. Particularly for Flat, the relative
abundance of the two species alternates across successive layers, giving
damped oscillations of the ionic charge density toward the interior. 
Step and Bump produce additional near-surface structure associated with their
different exposed heights. The oscillations for Step are less pronounced.

The lateral maps in \figrefsub{b}{fig:edl-structure} resolve the spatial
organization hidden by the planar averaging. The purple windows in \figrefsub{a}{fig:edl-structure} select
the first-layer region for each geometry. Over Flat, both species show
periodic density modulation imposed by the atomic lattice. The Step
organizes this modulation into an extended band, whereas the Bump
produces a localized patterned region. The geometrical imprint is
present in both species, with different populations, showing how
atomic-scale packing and charge separation respond together to the
surface feature. 
Overall, Equi-cDFT reproduces both the planar profiles and lateral maps of MD, 
although the MD maps have noticeable statistical noise.

\begin{figure*}[!t]
\centering
\includegraphics[width=\textwidth]{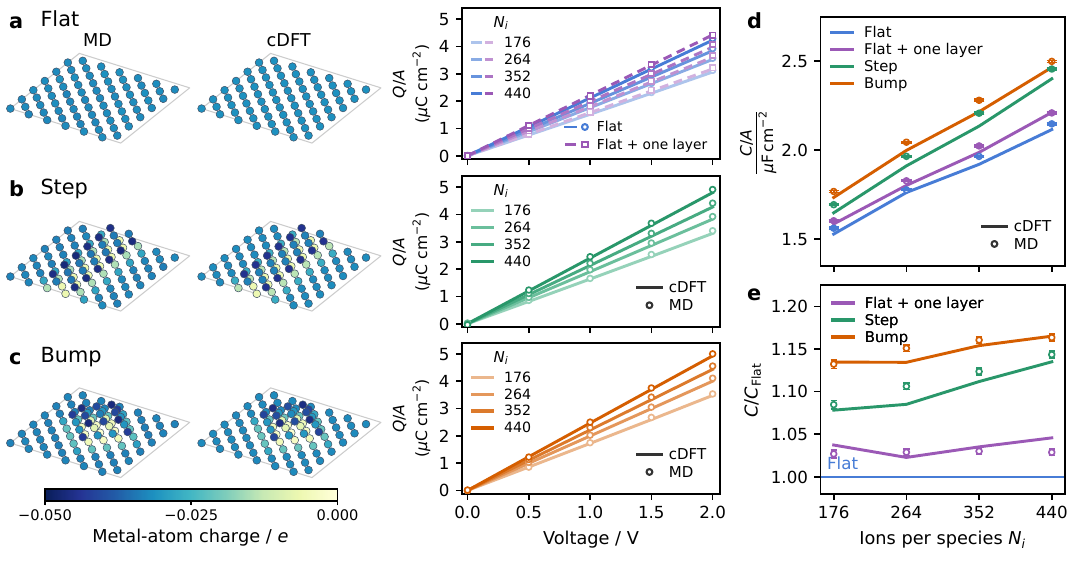}
\caption{\textbf{Geometry-dependent electrode charge and capacitance.}
\textbf{a--c} Flat, Step and Bump, respectively. Left: MD and cDFT
charges on the negative electrode surface,
at $N_+=N_-=352$ and $V=2\,\mathrm V$.
Right: electrode charge per projected area, $Q/A$,
for different numbers of ions of each species.
Panel \textbf{a} also includes the Flat plus one layer control, with one additional metal layer.
\textbf{d} Effective cell capacitance per projected area, $C/A$.
\textbf{e} Capacitance normalized by the Flat value at the same loading. MD error bars represent the standard error.}
\label{fig:edl-capacitance}
\end{figure*}

Maintaining a common metal potential requires different charges on atoms
in different local environments. The maps in
\figref{fig:edl-capacitance} show nearly equivalent surface-site charges
on Flat, whereas Step and Bump develop strongly charged exposed sites
alongside weakly charged regions of the base plane. The increase in
total electrode charge thus accompanies a heterogeneous redistribution
among metal atoms. Equi-cDFT captures both this site-resolved response
and the surrounding ionic structure in \figref{fig:edl-structure}
through their self-consistent coupling. A laterally averaged description
cannot resolve these coupled variations in ionic density and metal-site
charge.

Despite this microscopic heterogeneity, the charge--voltage curves are
approximately linear over the $0$--$2\,\mathrm V$ range studied.
Higher ionic loading supports larger electrode charge at the same
voltage, and geometry primarily changes the slope over this range.
To compare the slopes, we fit $Q/A=(C/A)V$ over $0$--$2\,\mathrm V$, where $C$ is a finite-window
effective cell capacitance.
\figrefsub{d}{fig:edl-capacitance} shows that Equi-cDFT captures both
the increase with loading and the ordering across geometries. Its
capacitances are typically about 2\% below the MD estimates.

Normalizing by Flat makes the effect of the surface features clearer
(\figrefsub{e}{fig:edl-capacitance}). Both the Step and the Bump enhance
charge storage, with Bump giving the larger enhancement throughout the
sampled range. The relative enhancement of Step grows with ionic
loading, narrowing its difference from Bump. The enhancement therefore
depends jointly on topography and ionic loading, rather than behaving
as a fixed geometrical multiplier.
Equi-cDFT captures this loading-dependent ordering. The laterally flat
added-layer control gives a smaller change that is nearly independent
of loading in MD. This contrast highlights the different responses to
a uniform interface displacement and to localized surface structure.

Pronounced planar layering is not an indicator of greater
charge storage. Flat exhibits clear oscillations extending into the
fluid, yet Step and Bump have larger capacitances. The weaker
planar-averaged oscillations for Step coexist with substantial lateral
organization near the raised strip. The lateral maps therefore reveal
interfacial structure that is obscured by planar averaging.

Atomistic studies have shown that surface corrugation alters capacitance
through changes in ionic
organization~\cite{Vatamanu2011Topography}, and that adsorption-site
geometry controls local charge storage in nanoporous
carbons~\cite{Merlet2013Confinement}. Surface patterning changes both how
much charge an electrode stores and how that charge is distributed among
its atoms. The loading-dependent enhancement shows that the benefit of
a surface feature must be assessed together with the surrounding ionic
fluid. For charge-storage applications, this motivates resolving the
heterogeneous local environments, and selecting surface
motifs for the intended electrolyte conditions.

\subsection{Charging dynamics}

Charge-storage capacity and charging rate are complementary measures of
electrode performance, with a possible
trade-off~\cite{Lian2020BlessingCurse}.
At a structured electrode, the ionic arrangements that support charge
storage must also reorganize during charging. We therefore ask whether
the same functional that describes the equilibrium interface can predict
this reorganization when combined with independently determined transport
coefficients.

\begin{figure*}[!t]
\centering
\includegraphics[width=\textwidth]{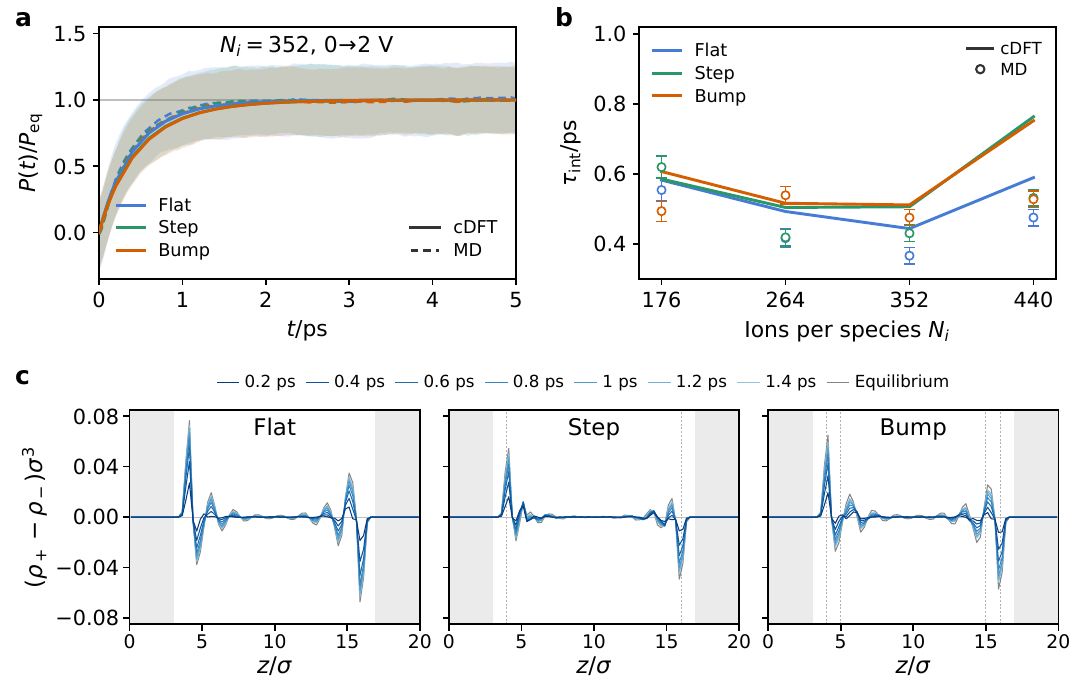}
\caption{\textbf{Charging dynamics at surface-structured electrodes.}
\textbf{a} Ionic polarization $P(t)/P_{\mathrm{eq}}$ at
$N_+=N_-=352$, where $P_{\mathrm{eq}}$ is each method's corresponding
$2\,\mathrm V$ equilibrium polarization reference.
Starting from an independently determined
$0\,\mathrm V$ equilibrium configuration,
the voltage is switched instantaneously to $2\,\mathrm V$.
Shading shows the MD sample standard deviation.
\textbf{b} Integral charging time
$\tau_{\mathrm{int}}$, with error bars showing MD standard error.
\textbf{c} Planar charge-number density
from cDFT at $N_+=N_-=352$.
Equilibrium profiles come from independent same-grid solves.
Shaded electrode regions and dotted feature-plane heights indicate the
geometry.}
\label{fig:edl-dynamics}
\end{figure*}

We perform dynamical cDFT calculations using Eq.~\eqref{eq:ddft_flux}, 
with the local mobility calibrated from independent bulk MD
(see Methods). 
We use the same three geometries and four ionic loadings as in the
equilibrium calculations, with a $0.25\sigma$ grid.
The initial density fields come from the equilibrium solves at $0\,\mathrm V$, and the voltage is instantaneously switched to $2\,\mathrm V$.
We then compare the dynamical cDFT with the mean of 1000 MD
trajectories following the same $0$ to $2\,\mathrm V$ switch.

At the displayed loading, the normalized polarization histories nearly
collapse across the three geometries
(\figrefsub{a}{fig:edl-dynamics}). Thus, the enhanced equilibrium charge
storage at Step and Bump is accompanied by broadly similar 
relaxation of the ionic polarization. Equi-cDFT captures this weak
geometry dependence and the main rise of the MD ensemble means,
although its initial relaxation is somewhat slower.
The broad MD bands show substantial trajectory-to-trajectory fluctuations.

The integral time in \figrefsub{b}{fig:edl-dynamics} characterizes the charging rate. Its loading dependence differs
from that of the equilibrium capacitance: capacitance increases with
particle number, whereas the integral times vary nonmonotonically,
with different trends across geometries. Charge-storage capacity alone
therefore does not determine the charging rate. 

The agreement between dynamical cDFT and MD is less uniform than the equilibrium one. 
At the highest loading, cDFT predicts a stronger slowdown than MD,
particularly for Step and Bump. 
The discrepancy probably stems from the approximations in the dynamics equation (Eq.~\eqref{eq:ddft_flux}) and the mobility matrix, particularly at the higher ionic density.

The time-resolved charge profiles in \figrefsub{c}{fig:edl-dynamics}
reveal how the interfacial screening structure develops. Charge
accumulates most strongly near the metal surfaces, while alternating
charge layers build up farther into the fluid as relaxation proceeds.
Step retains much weaker planar-averaged oscillations in the interior
than Flat and Bump. The similar normalized polarization histories
therefore arise from distinct spatial patterns of charge redistribution:
a single integrated charging curve does not reveal where the ionic
response occurs.

\section{Discussion}

We extend Equi-cDFT to ionic fluids by combining species-resolved,
symmetry-adapted short-range density descriptors with explicit long-range
electrostatics. 
Minimizing the same scalar free-energy functional gives equilibrium ionic
structures around charged solutes and structured electrodes. Combining
it with mobilities determined independently from bulk MD also describes
electrode charging dynamics.

Bui and Cox combined neural functionals for short-range mimic fluids
with local molecular field theory to describe ionic
fluids~\cite{Bui2025Ionic}, and subsequently applied related ideas to
field-driven restructuring of polar fluids~\cite{bui2026dielectrocapillarity}.
Here we learn the scalar free energy
directly from full-Coulomb simulations and use it to describe ionic
structure in three dimensions.
It is also worth noting that, for species with environment-dependent charges, 
one might be able to learn the long-range Coulomb interactions using
latent Ewald Summation type schemes~\cite{Cheng2025Latent}.

Furthermore, for modeling EDLs, the learned fluid functional is coupled to explicit metal atoms whose charges adjust to maintain the imposed
potential. Molecular DFT has previously been employed together with atomistic
constant-potential electrodes~\cite{Jeanmairet2019WaterGraphene}.
This coupling resolves how ionic packing around surface features and
the redistribution of metal charge influence one another.

The present RPM functional can be applied to more complex geometries,
including nanoporous electrodes~\cite{Merlet2013Confinement}, assemblies
of charge-patterned cubes or rods, and charged particles near surfaces.
A further direction is to learn functionals for more realistic ionic
fluids, such as aqueous electrolytes, mixed molten salts, and organic electrolyte solutions.

\section{Methods}

\subsection{RPM model}

The restricted primitive model consists of two equal-diameter ionic species
with charges $q_a=z_ae$, where $z_+=1$ and $z_-=-1$, in a uniform
dielectric continuum. We use $\sigma=2.76\,\text{\AA}$,
$T=4000\,\mathrm K$, and $\epsilon_r=1$. Length, species density and
potential energy are expressed as $r^*=r/\sigma$,
$\rho_a^*=\rho_a\sigma^3$ and $V_a^*=V_a/E_C$, where
$E_C=e^2/(4\pi\epsilon_0\sigma)\approx5.217\,\mathrm{eV}$ is the
Coulomb energy scale.
Thus $T^*=k_{\mathrm B}T/E_C\simeq0.066$ and
$\beta V_a=V_a^*/T^*$.

In the MD simulations, ions interact through the pair potential
$u_{ab}(r)=u_{\mathrm c}(r)+E_Cz_az_b\sigma/r$. The same purely
repulsive core is used for all species pairs: a 50--49 Mie potential
truncated and shifted at its minimum~\cite{Bui2025Ionic},
\begin{equation}
u_{\mathrm c}(r)=
\begin{cases}
\varepsilon_{\mathrm c}\!\left[A_{50,49}(x^{50}-x^{49})+1\right],
&r<r_{\mathrm{ion}},\\
0,&r\ge r_{\mathrm{ion}},
\end{cases}
\label{eq:md_ion_core}
\end{equation}
where $x=\sigma/r$, $A_{50,49}=50(50/49)^{49}$,
$r_{\mathrm{ion}}=50\sigma/49$ and
$\varepsilon_{\mathrm c}=k_{\mathrm B}T/1.5$.
This continuous pseudo-hard-sphere core approximates excluded volume.

MD simulations use LAMMPS~\cite{Thompson2022LAMMPS} in the canonical
ensemble, with cation and anion masses of 22.990 and
$35.45\,\mathrm{g\,mol^{-1}}$, respectively. Ionic trajectories use
velocity-Verlet integration with a $1\,\mathrm{fs}$ production timestep
and a Langevin thermostat with damping time $100\,\mathrm{fs}$;
the net random thermostat force is removed. Coulomb interactions use a
$10\,\text{\AA}$ real-space cutoff, with the remaining interaction
evaluated by particle--particle particle--mesh (PPPM) Ewald summation.
The requested PPPM accuracies are $10^{-5}$ for bulk and solute-particle
simulations and $10^{-6}$ for electrode simulations.

\subsection{Equi-cDFT model details}

The production model resolves the two RPM number densities on a training
grid with $h_0=0.25\sigma$. The architecture uses a physical cutoff
$r_{\mathrm{cut}}=1.5\sigma$ and two normalized, origin-centered Gaussian radial
functions with trainable widths. Cartesian moments have total degree
$\ell_x+\ell_y+\ell_z\le3$, with invariant products of order $\nu\le2$.
A nonnegative nearest-neighbor average of width $\ell_{\mathrm{sm}}=h_0/\sqrt6$ smooths the densities
entering the learned branch.
The ideal, external-potential and analytical Coulomb terms use the
original densities. 
The Coulomb term in Eq.~\eqref{eq:long_range_energy}
uses $\alpha=0.20\sigma^2$.
The invariant features, central densities and temperature use fixed
reference scales and feed a fully connected network with two hidden
layers of 32 and 16 neurons, SiLU activations and linear species-resolved
outputs. The readout couples species and radial channels, giving 2612
trainable parameters including the radial widths.

The cDFT applications used a single NVIDIA L40S or RTX 6000 Ada GPU.
Evaluating the excess free energy and its first density derivatives on a
$48\times48\times80$ grid took approximately $50\,\mathrm{ms}$.

\subsection{Training set}

Training uses canonical density fields of the full-Coulomb RPM under
species-dependent external potentials in periodic cubic cells of side
$8\sigma$, with $N_+=N_-$ in each cell. Box-averaged per-species reduced
densities span approximately $0.025\le\bar\rho_i\sigma^3\le0.4$.
The dataset contains 1570 training and 139 validation configurations.

The external potentials are random superpositions of periodically wrapped
Gaussian wells and barriers with one-, two- and three-dimensional spatial
dependence. Individual components have widths of $0.2$--$3\sigma$ and
amplitude magnitudes of $0.2$--$6k_{\mathrm B}T$. The fields are applied
identically to both species ($V_+=V_-$), with opposite signs
($V_+=-V_-$), or as independently drawn species fields. Other configurations
combine common repulsive fields with weaker charge-separating perturbations,
or include neutral repulsive slit walls.

Species-resolved density histograms are accumulated every five MD steps
($5\,\mathrm{fs}$) and averaged over $16$--$20\,\mathrm{ns}$.
The training fields use an $h_0=0.25\sigma$ grid with $32^3$ cells, with the
analytical external potentials evaluated at cell centers.
For training augmentation, the two species' density and external-potential
fields are exchanged together with probability $1/2$ per configuration
per epoch.

Bulk number--number and charge--charge structure factors, $S_{NN}(k)$
and $S_{ZZ}(k)$, from MD are species-symmetrized and used as additional
training targets. They cover 16 per-species densities
$\rho_i\sigma^3=0.025,0.050,\ldots,0.400$.

\subsection{Training}

Training combines a field loss based on the chemical-potential balance in
Eq.~\eqref{eq:chemical_potential_balance_loss} with a bulk-response loss
$\mathcal L_S$ that matches the predicted inverse number--number and
charge--charge structure factors, $1/S_{NN}(k)$ and $1/S_{ZZ}(k)$,
to their species-symmetrized MD counterparts. The field loss is evaluated
on unmasked voxels with $\rho_a^*>10^{-4}$.
Training uses Adam for 300 epochs. Epochs 1--100 use
the field loss alone.
During epochs 201--300, an additional Fourier-stability regularizer
penalizes negative curvature of the functional at equilibrium fields, evaluated using particle-number-conserving plane-wave perturbations.
We note that this regularizer has neglegible effect on the training,
and could be omitted.

On the 139 validation configurations, forward density-to-potential
predictions give an RMSE of $0.158\,k_{\mathrm B}T$ and
$R^2=0.9894$, after aligning the additive potential constant for each
species. 
Reverse finite-iteration minimization from uniform densities give an
RMSE of $8.14\times10^{-3}\sigma^{-3}$ and $R^2=0.9950$.

\subsection{Grid-spacing transfer}

Grid transfer evaluates the descriptors at the new voxel centers while
preserving their physical cutoff, radial functions and smoothing width.
Integration weights are adjusted to the new voxel volume, with all
learned parameters unchanged.

\begin{figure}[t]
\centering
\includegraphics[width=\columnwidth]{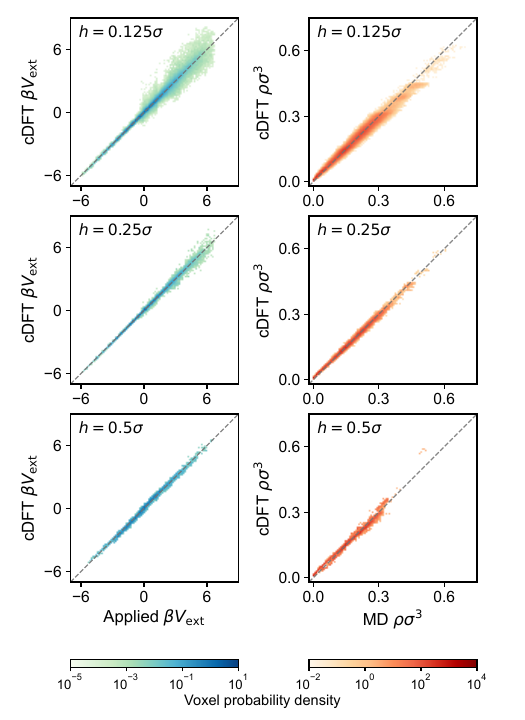}
\caption{\textbf{Grid-spacing transfer without retraining.}
The left column compares reconstructed and applied external potentials;
the right compares finite-iteration cDFT density predictions with MD.
Rows correspond to $h/\sigma=0.125$, $0.25$ (the training grid, $h=h_0$), and $0.5$.}
\label{fig:grid-transfer}
\end{figure}

We evaluate the unchanged production functional on grids with
$h/\sigma=0.125$, $0.25$, and $0.5$. The comparison in \figref{fig:grid-transfer}
uses three existing validation configurations in a periodic $8\sigma$
cube with 77 ions of each species: common fields ($V_+=V_-$), charge
fields ($V_+=-V_-$), and independently drawn species fields. Each parity
plot pools both ionic species across the three configurations.
The MD densities on the three grids are sampled over 20 ns, and the
analytical external potentials are evaluated at the voxel centers.
For the forward test, we reconstruct $\beta V^{\mathrm{ext}}$ from the
MD densities using the Euler--Lagrange relation, aligning one additive
constant per species on voxels with $\rho_a\sigma^3>10^{-4}$.
For the reverse test, we start from spatially uniform densities at the
same particle numbers and run Euler iterations. 
All species and voxels enter the density parity plots. 
MD voxel noise also changes with grid resolution.

\subsection{EOS and charge correlation}
\label{sec:bulk-validation}

The bulk MD references contained 2000 ions of each species in periodic cubes
at 16 per-species densities, $\rho_i\sigma^3=0.025,0.050,\ldots,0.400$.
Each state contributed $2\,\mathrm{ns}$ of production, from which we obtained
the kinetic-plus-virial pressure. For the four structure-factor comparisons at
$\rho_i\sigma^3=0.1,0.2,0.3,0.4$, we evaluated
Eq.~\eqref{eq:charge_structure_factor} from ion coordinates and averaged
over reciprocal shells in $0<k\sigma\le3.14$.

\begin{figure}[t]
\centering
\includegraphics[width=\columnwidth]{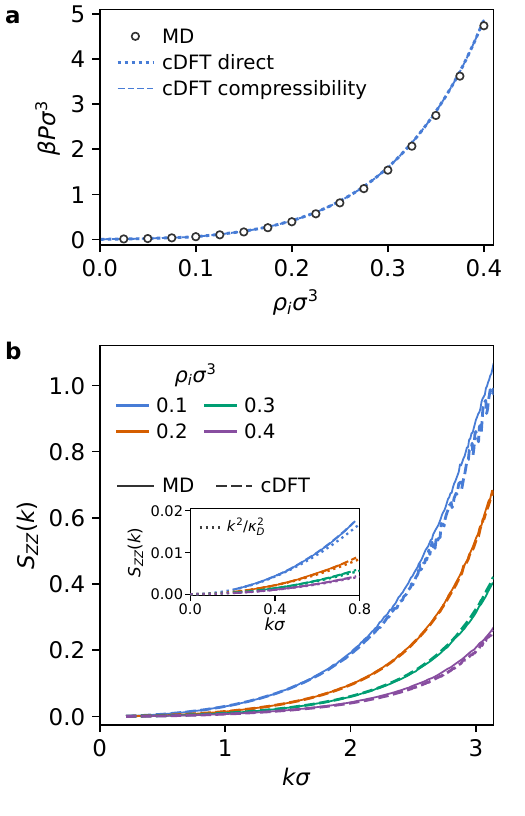}
\caption{\textbf{Bulk thermodynamics and charge correlations of the RPM.}
\textbf{a} Equation of state from MD and the direct and compressibility
cDFT routes. \textbf{b} $S_{ZZ}(k)$ from MD and cDFT. The inset
compares the low-$k$ behavior with the Stillinger--Lovett limit.
Error bars and shading indicate MD standard error.}
\label{fig:bulk-response-phase}
\end{figure}

For homogeneous neutral fluids, $\rho_i=\rho_+=\rho_-=\rho/2$ with total
number density $\rho=N_{\mathrm{tot}}/V$, the Helmholtz free-energy density
$f_T(\rho)=(F_{\mathrm{id}}+F_{\mathrm{exc}})/V$ gives the direct pressure
\begin{equation}
P(\rho,T)=\rho\frac{\partial f_T}{\partial\rho}-f_T(\rho).
\label{eq:results_direct_pressure}
\end{equation}
For the compressibility route, we differentiate
$c^{(1)}=(c_+^{(1)}+c_-^{(1)})/2$ along the
same neutral path:
\begin{equation}
c^{(2)}(\mathbf r_g,\mathbf r_{g'};[\rho],T)
=\frac{1}{\Delta V}
\frac{\partial c^{(1)}(\mathbf r_g;[\rho],T)}
{\partial \rho(\mathbf r_{g'})}.
\label{eq:learned_second_direct_correlation}
\end{equation}
Using $\widehat c^{(2)}(0)=\Delta V\sum_g c^{(2)}(\mathbf r_g)$ and
$\partial(\beta P)/\partial\rho=1-\rho\widehat c^{(2)}(0;\rho,T)$,
integration from the ideal-gas limit gives
\begin{equation}
\beta P(\rho,T)=\int_0^\rho\!\mathrm d\rho'\,
\left[1-\rho'\widehat c^{(2)}(0;\rho',T)\right].
\label{eq:results_compressibility_pressure}
\end{equation}
The two routes agree and closely follow the independent MD pressure reference
(\figrefsub{a}{fig:bulk-response-phase}). Their mutual agreement checks
consistency between the scalar free energy and its derivatives.

Charge correlations are measured by
\begin{equation}
S_{ZZ}(k)=\frac{1}{N_{\mathrm{tot}}}
\left\langle\left|\sum_j z_j
e^{-i\mathbf k\cdot\mathbf r_j}\right|^2\right\rangle,
\label{eq:charge_structure_factor}
\end{equation}
where $z_j=\pm1$.
The cDFT spectra are obtained from the mixture direct correlations in a
$30\sigma$ cubic box with grid spacing $0.25\sigma$.
They reproduce the density dependence of MD and approach
$S_{ZZ}(k)\simeq k^2/\kappa_D^2$ at low $k$
(\figrefsub{b}{fig:bulk-response-phase}). Here
$\kappa_D^2=8\pi\ell_{\mathrm B}\rho_i$, with Bjerrum length
$\ell_{\mathrm B}=e^2/(4\pi\epsilon_0\epsilon_r k_{\mathrm B}T)$.

\subsection{Mobility from bulk density correlations}

To derive Eq.~\eqref{eq:mobility_from_md}, consider fluctuations about a
homogeneous equilibrium state with uniform density $\boldsymbol\rho_0$.
The thermodynamic restoring force is set by
the Hessian of the total free energy,
\begin{equation}
H_{ab}(\mathbf r-\mathbf r')
=\left.\frac{\delta^2F}
{\delta\rho_a(\mathbf r)\delta\rho_b(\mathbf r')}\right|_{\boldsymbol\rho_0}.
\label{eq:mobility_hessian}
\end{equation}
The first-order change in chemical potential is
\begin{equation}
\delta\mu_a(\mathbf r,t)=\sum_b\int_V
H_{ab}(\mathbf r-\mathbf r')\delta\rho_b(\mathbf r',t)\,\mathrm d\mathbf r'.
\end{equation}
The spatial Fourier transform for $k\ne0$ gives
\begin{equation}
    \widetilde{\rho}_{a}(\mathbf k,t)
    =\dfrac{1}{V}\int_V d \mathbf r  \delta\rho_a(\mathbf r, t) e^{-i\mathbf k \cdot \mathbf r}
    =\dfrac{1}{V}\sum_{j\in a}e^{-i\mathbf k\cdot\mathbf r_j(t)},
\end{equation}
\begin{equation}
\mathbf H(\mathbf k)=\int_V
\mathbf H(\mathbf r)e^{-i\mathbf k\cdot\mathbf r}\,\mathrm d\mathbf r,
\end{equation}
and
\begin{equation}
\delta \widetilde{\boldsymbol\mu} (\mathbf k, t) =\mathbf H(k)
\widetilde{\boldsymbol\rho}(\mathbf k, t).
\end{equation}
Equilibrium fluctuation--response thus gives~\cite{Evans1979}
\begin{equation}
\left\langle
\widetilde{\boldsymbol\rho}(\mathbf k,t)
\widetilde{\boldsymbol\rho}^{\dagger}(\mathbf k,t)
\right\rangle
= \mathbf{C}(\mathbf{k},0)
= \dfrac{k_{\mathrm B}T}{V} \mathbf H(k)^{-1},
\label{eq:Ck0}
\end{equation}
where $\dagger$ denotes the conjugate transpose.

Let $\widetilde j_a=(\mathbf k/k)\cdot\widetilde{\mathbf J}_a$
be the longitudinal flux of species $a$, with chemical-potential and flux
modes using the same volume-normalized transform as the density modes.
Density conservation in dynamic evolution gives
\begin{equation}
\partial_t \widetilde{\boldsymbol\rho}(\mathbf k, t)
=
-ik \widetilde{\boldsymbol j}(\mathbf k,t),
\label{eq:conservation}
\end{equation}
where the longitudinal flux can be expressed as
\begin{equation}
\widetilde{\boldsymbol j}(\mathbf k,t)
=-ik\int_0^t\mathbf K(k,\tau)
\delta \widetilde{\boldsymbol \mu} (\mathbf k, t-\tau)\,\mathrm d\tau
+\widetilde{\boldsymbol j}^{R}(\mathbf k,t).
\label{eq:mobility_memory_flux}
\end{equation}
The projection~\cite{Mori1965Transport} onto the density modes represents
the unresolved degrees of freedom by a causal memory kernel $\mathbf K(k,\tau)$
and a random remainder orthogonal to the initial density,
$\langle\widetilde{\boldsymbol j}^{R}(\mathbf k,t)
\widetilde{\boldsymbol\rho}^{\dagger}(\mathbf k,0)\rangle=\mathbf0$.
Right-multiplying Eq.~\eqref{eq:conservation} by
$\widetilde{\boldsymbol\rho}^{\dagger}(\mathbf k,0)$ and averaging yields
\begin{equation}
\partial_t\mathbf C(k,t)
=-k^2\int_0^t\mathbf K(k,\tau)\,\mathbf H(k)\,
\mathbf C(k,t-\tau)\,\mathrm d\tau.
\label{eq:mobility_time_regression}
\end{equation}

Define the time-integrated mobility
\begin{equation}
\mathbf M(k)=\int_0^\infty\mathbf K(k,\tau)\,\mathrm d\tau.
\end{equation}
With $\mathbf I(k)$ defined in Eq.~\eqref{eq:Ik}, assume at fixed nonzero
$k$ that the correlations decay to zero, the time integrals converge, and
$\mathbf I(k)$ is invertible. Integrating
Eq.~\eqref{eq:mobility_time_regression} and interchanging the order of
time integration gives
\begin{equation}
-\mathbf C(k,0)=-k^2\mathbf M(k)\,\mathbf H(k)\,\mathbf I(k).
\end{equation}
Right-multiplying by $\mathbf I(k)^{-1}\mathbf H(k)^{-1}$ and using
Eq.~\eqref{eq:Ck0} yields~\cite{Akaberian2023Mobility}
\begin{equation}
\mathbf M(k)=\frac{V}{k_{\mathrm B}T\,k^2}
\mathbf C(k,0)\,\mathbf I(k)^{-1}\,\mathbf C(k,0).
\label{eq:mobility_integrated_regression}
\end{equation}
The time integrals are evaluated at fixed nonzero $k$, before taking the
long-wavelength limit.

\subsection{Mobility calibration from bulk MD}

The mobility depends on the Langevin thermostat used in the reference MD.
The calibration uses two independent 16 ns replicas containing 9,826 ions
of each species in periodic cubes at each per-species density
$\rho_i\sigma^3=0.1,0.2,0.3,0.4$. Species-density correlation matrices are
computed in reciprocal space and averaged over equivalent wavevectors.

\begin{figure}[t]
\centering
\includegraphics[width=\columnwidth]{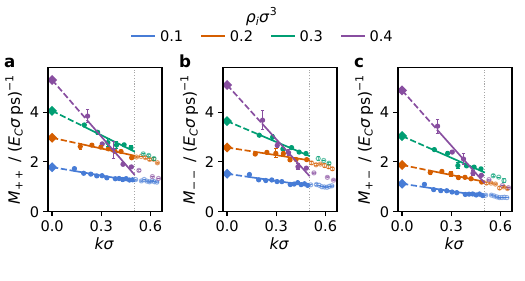}
\caption{\textbf{Species-resolved mobility and small-$k$ extrapolation.}
\textbf{a} $M_{++}$, \textbf{b} $M_{--}$, and
\textbf{c} $M_{+-}=M_{-+}$.
Results are shown at per-species densities $\rho_i\sigma^3=0.1,0.2,0.3,0.4$.
Filled and open circles denote points inside and outside the fitting
range; dashed extensions and diamonds show the extrapolation to $k=0$.
Error bars show standard error.}
\label{fig:mobility-extrapolation}
\end{figure}

We integrate the correlations up to $t_c=100\,\mathrm{ps}$ and obtain
the finite-window mobility $\mathbf M(k;t_c)$ from
Eq.~\eqref{eq:mobility_from_md}, averaging replica static matrices and
correlation integrals before inversion. Unweighted linear fits
$\mathbf M(k;t_c)=\mathbf A+\mathbf D(k\sigma)$ over
$0<k\sigma\le0.5$ give the long-wavelength estimates $\mathbf M^0$
as the intercepts (\figref{fig:mobility-extrapolation}).

Let $\mathbf B(\rho)$ denote the piecewise-linear interpolation of the four
bulk mobility matrices as a function of per-species density $\rho$.
The local mobility used in Eq.~\eqref{eq:ddft_flux} is
\begin{equation}
\begin{aligned}
M^0_{ab}(\{\rho\})
&=\sqrt{\rho_a\rho_b}\,
\frac{B_{ab}(\rho_{\mathrm{clip}})}{\rho_{\mathrm{clip}}},\\
\rho_{\mathrm{clip}}&=\operatorname{clip}\!\left(
\frac{\rho_++\rho_-}{2},0.1\sigma^{-3},0.4\sigma^{-3}\right).
\end{aligned}
\label{eq:local_mobility}
\end{equation}
Outside the measured interval the mobility per density is held at the
nearest endpoint. 

\subsection{Reference MD simulations for charged particles}
\label{sec:md-charged-particles}

Simulations of the solute particles used a periodic $(16\sigma)^3$ cell with
$N_+=N_-=820$, with both solute centers and orientations held fixed.
The prolate ellipsoids carried central charges $+e$ and $-e$; each
neutral bipolar dumbbell carried $+e$ and $-e$ at its lobe centers.
For each orientation, 15 surface gaps
spanned approximately $0.108\sigma$ to $3.024\sigma$ at intervals
$5\sigma/24$. 
A surface gap is the distance
between the opposing $10k_{\mathrm B}T$ core contours along the
center-separation axis.
Each configuration contributed two separately seeded $10\,\mathrm{ns}$
production replicas. 

Each solute is represented by a soft repulsive core and fixed point
charges. The core is constructed from the radial potential
\begin{equation}
u_0(r)=k_{\mathrm B}T\left[
\frac{2}{15}\left(\frac{2\sigma}{r}\right)^9
-\left(\frac{2\sigma}{r}\right)^3-C_{\mathrm{core}}
\right],
\label{eq:md_colloid_core}
\end{equation}
for $r<r_{\mathrm{solute}}$ and is zero otherwise. Here
$r_{\mathrm{solute}}=2\sigma(2/5)^{1/6}$, and $C_{\mathrm{core}}$ makes
the potential zero at the cutoff. We define the nominal particle surface by the
$u_0=10k_{\mathrm B}T$ contour, whose radius for the spherical
reference is $R_0\simeq1.196\sigma$. Both particle shapes have the
same volume enclosed by this contour.

For the prolate particles, the repulsion is
$u_0(r_{\mathrm{eff}})$, with
\begin{equation}
r_{\mathrm{eff}}
=R_0\sqrt{\frac{d_\parallel^2}{a^2}
+\frac{d_\perp^2}{b^2}},
\end{equation}
where $d_\parallel$ and $d_\perp$ are the components of the
minimum-image ion--center displacement parallel and perpendicular
to the particle axis. The semiaxes
$a=2^{2/3}R_0$ and $b=c=2^{-1/3}R_0$ give an aspect ratio of two
while preserving the reference volume.

Each dumbbell consists of two overlapping spherical lobes with
nominal radius $R_d\simeq0.963\sigma$ and center separation $1.5R_d$.
The repulsion of each lobe is $u_0(R_0d/R_d)$, where $d$ is the
distance from its center. The dumbbell potential is the maximum
of the two lobe potentials.


\subsection{cDFT for charged particles}

The equilibrium cDFT calculations use the same solute geometries, orientations and gap range. We use
L-BFGS minimization of the production functional in a periodic
$(16\sigma)^3$ cell on a $0.125\sigma$ grid, with fixed ionic particle
numbers $N_+=N_-=820$. For each orientation, the cDFT force curve uses
29 uniformly spaced surface gaps, with spacing $5\sigma/48$.

For the force integral in Eq.~\eqref{eq:colloid_feynman_hellmann_grid},
midpoint quadrature is used except in voxels crossing the lobe-join planes
of the axial dumbbells, where discontinuities in the potential gradient
are treated with Boltzmann-weighted subvoxel quadrature.

Equilibrium calculations typically took less than a minute on the
$0.25\sigma$ grid and approximately seven minutes on the $0.125\sigma$
grid per configuration.

\subsection{Reference MD simulations for EDLs}
\label{sec:md-edls}

The electrode cell was periodic with dimensions
$12\sigma\times12\sigma\times20\sigma$. Fixed FCC(100) sites with
lattice constant $2\sigma$ formed a metal slab, exposing two surfaces to the fluid. The flat geometry
contained 504 sites, with exposed planes at $z/\sigma=3$ and 17.
The narrow step added 24 sites on each surface at $z/\sigma=4$ and 16,
selected by $4\le x/\sigma<8$, giving 552 sites. The bump added layers
within radii $3\sigma$ and $2\sigma$ about $(x,y)=(6,6)\sigma$ at
$z/\sigma=4$ and 5, reflected to 16 and 15, giving 554 sites.
A laterally flat control added a complete 72-site lower layer at
$z/\sigma=4$, giving 576 sites while retaining the original upper surface.
All metal atoms remained frozen at their lattice positions throughout
the equilibrium and charging simulations; only their charges could change.

We used the Siepmann--Sprik fluctuating-charge metal
model~\cite{siepmann1995influence}, with an explicit Gaussian charge on
each electrode atom and inverse width $\eta=1/(0.2\sigma)$.
An ion of species $a$ interacted with a metal atom $j$ through
\begin{equation}
\begin{split}
u^{\mathrm{im}}_{aj}(r)={}&4\varepsilon_{\mathrm{im}}S(r)
\left[\left(\frac{\sigma_{\mathrm{im}}}{r}\right)^{12}
-\left(\frac{\sigma_{\mathrm{im}}}{r}\right)^6\right]\\
&+\frac{z_a e Q_j}{4\pi\epsilon_0\epsilon_r}
\frac{\operatorname{erf}(\eta r)}{r},
\end{split}
\label{eq:ion_metal_interaction}
\end{equation}
where $r$ is the ion--metal separation and $Q_j$ is the instantaneous
charge on metal atom $j$. The first term describes excluded volume and
short-range attraction, with $\sigma_{\mathrm{im}}=1.25\sigma$ and
$\varepsilon_{\mathrm{im}}=0.1k_{\mathrm B}T$ for both ionic species.
The switching function is $S(r)=1$ below $2.5\sigma$, and switches smoothly to zero at $3.125\sigma$.

The metal charges were relaxed at every ionic timestep using the
matrix-inversion constant-potential implementation in the LAMMPS
ELECTRODE package~\cite{AhrensIwers2021Mesh,AhrensIwers2022Electrode}.
The total metal charge was constrained to zero. Finite-field mode imposed
$E_z=-V/L_z$, where $V$ denotes the voltage across the full periodic cell.

For equilibrium voltage scans we used $N_+=N_-=176,264,352,440$ and
$V=0,0.5,1,1.5,2\,\mathrm V$. Each state had two separately
seeded replicas with $2\,\mathrm{ns}$ of production each.
Densities and charge observables were sampled every $0.1\,\mathrm{ps}$,
using $0.125\sigma$ and $0.25\sigma$ density grids.

Charging references used 1000 separately seeded $10\,\mathrm{ps}$ trajectories for each geometry and loading. Starting from zero-voltage
equilibrium states, we switched instantaneously to $2\,\mathrm V$.

The ionic polarization is
$P(t)=V^{-1}\sum_{i\in\mathrm{ions}}q_i z_i(t)$. MD and cDFT histories are normalized
by their respective $2\,\mathrm V$ equilibrium polarizations. For MD
trajectory $j$, the charging time is
\begin{equation}
\tau_j=\int_0^{10\,\mathrm{ps}}
\left[1-\frac{P_j(t)}{P_{\mathrm{eq}}}\right]\,\mathrm dt.
\label{eq:md_charging_time}
\end{equation}
The MD value of $\tau_{\mathrm{int}}$ is the mean of the 1000 individual
$\tau_j$. 

\subsection{cDFT for EDLs}

The equilibrium cDFT calculations use the same cell, fixed metal sites,
ion--metal interactions, particle numbers and voltages as the MD
reference. We solve for the ionic densities on a $0.125\sigma$ grid at
fixed species particle numbers using L-BFGS minimization.
For each trial ionic density, the Gaussian metal-site charges minimize
their periodic electrostatic energy, including ion--metal coupling,
metal--metal interactions, Gaussian self-energies and the applied-field
work. This quadratic problem is solved as a linear system subject to
zero total metal charge, treating the connected slab as one neutral
conductor. The relaxed charges enter the free energy and its density
derivatives self-consistently at the imposed voltage.

For the displayed lateral maps, densities are averaged over fixed
$0.5\sigma$-thick first-layer windows. The lower windows are
$z/\sigma\in[4,4.5)$ for Flat, $[5,5.5)$ for Step and $[6,6.5)$ for
Bump; the upper windows are their reflections about $z=10\sigma$.
The same windows are used across loadings and voltages. For both MD
and cDFT, the upper-layer map is reflected and its species labels
exchanged before averaging with the lower-layer map; only the resulting
lower representative is displayed. 

Equilibrium calculations typically took tens of seconds on the
$0.25\sigma$ grid and a few minutes on the $0.125\sigma$ grid.

\subsection{Dynamical cDFT simulation setup}

We simulate the same geometries as the
equilibrium calculations. Each simulation starts from the corresponding
$0\,\mathrm V$ equilibrium density fields; the voltage is switched
instantaneously to $2\,\mathrm V$ at $t=0$, and the densities are evolved
for $10\,\mathrm{ps}$.
We integrate Eq.~\eqref{eq:ddft_flux} on a $0.25\sigma$
grid using an adaptive, flux-limited explicit Euler scheme that preserves
positivity and each species' particle number.
A $10\,\mathrm{ps}$ charging trajectory on the $0.25\sigma$ grid took
approximately five minutes.

\section*{Data availability}
The trained model, numerical source data, evaluation and application scripts,
and evaluation and reference datasets will be made publicly available at
\url{https://github.com/BingqingCheng/equicdft-rpm-data}.
The training and validation density fields will be distributed separately due to their larger sizes.

\section*{Code availability}
The Equi-cDFT library is publicly available at
\url{https://github.com/BingqingCheng/equicdft}.

\section*{Acknowledgements}

OpenAI Codex assisted with literature organization, manuscript editing, and the development and review of plotting and figure-generation scripts. 
Claude Code assisted with code restructing and reviewing.
The author directed its use and independently verified all scientific claims, calculations, code, figures, and references.

\section*{Author contributions}
BC designed and performed the study and wrote the paper.

\section*{Competing interests}
The author declares no competing interests.

\end{document}